\documentstyle[aps,epsf,preprint]{revtex}
\input {amssym.def}
\input {amssym.tex}
\everymath{\rm}
\everydisplay{\rm}

\begin{document}
\title{Metal-Insulator Transition of Disordered \\
Interacting Electrons}

\author{Qimiao Si}
\address{Physics Department, Rice University, Houston, TX 77251-1892}
\author{C. M. Varma}
\address{Bell Laboratories, Lucent Technologies \\
Murray Hill, NJ 07974}

\maketitle

\begin{abstract}
We calculate the corrections to the conductivity and compressibility 
of a disordered metal when the mean free path is smaller than 
the screening length.
Such a condition is  shown to be realized for low densities and
large disorder. Analysis of the stability of
the metallic state reveals a transition to the insulating state
in two-dimensions.
\end{abstract}

\newpage
The discovery of a metal-insulator transition in the two-dimensional
electron gas,in a Si-MOSFET\cite{Kravchenko1,Kravchenko2}
and subsequently observed in other
systems\cite{Popovic,Hanien,Simmons,Coleridge}, suggests that
there remains  much to be understood in this classic problem.
The available theory ignoring interactions predicts an insulating
state for any disorder at all densities\cite{Gang4}.
The most systematic theory, including effects of both disorder
and interactions, due to Finkelstein\cite{Finkelstein},
followed the discovery of singularities in the problem 
by Altshuler and Aronov\cite{Altshuler}.
It predicts a metallic state at all densities. 
The new experiments have also generated much recent theoretical
activity\cite{recent}.

Finkelstein's theory has, however, a remarkable prediction with
which experiments are consistent.  For a
magnetic field coupling to spins an insulating state
appears to occur at all densities at low enough
temperature\cite{Kravchenko2}.

Finkelstein's theory is based on the existence of two scaling 
variables - effective disorder, parametrized by the conductance,
$g$ and an effective dimensionless spin-spin interaction
parameter $\gamma_t$. The reduction to these two 
parameters is largely based on the assumption 
that compressibility must be
continuous across a metal-insulator transition.  This ensures that
the electron-electron interactions $\gamma_s$ in the singlet channel is
irrelevant.  Indeed, existing explicit
calculations on the metallic side show no singular correction to
the compressibility in leading order in 
disorder\cite{Finkelstein,Altshuler,Castellani}.

Here we seek to add an important physical feature to the
theory of interacting disordered fermions
so that the modified theory has a metal-insulator transition.
It might be argued that Finkelstein's theory scales 
at low temperatures to a
metallic state but with strong coupling in the spin-spin
interaction channel where the analysis breaks down.  
Could a strong-coupling analysis of the same theory 
lead to a new low energy scale below which an insulating state emerges? 
We believe the experimentally observed metal-insulator 
transition is not due to the emergence of a new energy
scale primarily because such a transition is apparent already
at temperatures of 
$O(E_F)$.\cite{Kravchenko1,Kravchenko2,Popovic,Hanien,Simmons,Coleridge}

We have been motivated to re-examine the question of
the renormalization of the compressibility by two arguments: 
First, the metal-insulator transition in the pure limit, i.e. the Wigner
transition.  As $r_s$ is increased, either in two or three
dimensions, a first-order transition to the Wigner crystal 
is expected to occur
due to the long-range nature of the Coulomb interaction even for
spin-less electrons. The Wigner transition appears to satisfy conditions
in which disorder turns a first order transition to a continuous
transition\cite{Imry}.
It seems surprising then that the singlet interaction
would become irrelevant in the disordered problem.  
Second, if the insulating state has unscreened interactions, as commonly
assumed, for instance in the Efros-Shklovskii\cite{Efros,Chandra}
derivation, the screening length on the metallic side 
must diverge as the metal-insulator transition is approached.
The screening length is generally proportional to the 
inverse compressibility, (see Eq. (\ref{screening.length}) below). 

The correction to the compressibility can be calculated
from the correction to the exchange and correlation
contribution to the ground state energy (per unit area) 
from the so-called ring-diagrams, with disorder, shown in Fig. (1):
\begin{eqnarray}
E_{ring}
=  \frac{i}{2} \int d \omega
\int d^2 q \ell n \left[ 1 + U (q) \pi (q , \omega ) \right] \:.
\label{Ering}
\end{eqnarray}
Here $U (q) = 2 \pi e^2 / \epsilon q$ in two-dimensions.
The {\it proper} polarization in the diffusive regime is 
\begin{eqnarray}
\pi \left( q, \omega \right) = \nu
\frac{Dq^2}{i \omega + Dq^2}, \:\mbox{for}\:  q << \ell^{-1}, 
\omega << \tau^{-1} . 
\label{polarization}
\end{eqnarray}
For other $q$ and $\omega$ which we refer to as the ballistic regime, 
the polarizability is given by the generalization of
the usual form\cite{Stern} 
to include the leading order contribution of impurity
scattering, $\omega \rightarrow \omega + i / \tau$
where $1/\tau$ is the single-particle scattering
rate\cite{impurity}.
In Eq. (\ref{polarization}) D is the diffusion constant, 
$\ell$ the mean free path and $\nu$ the density of states. 
The compressibility at fixed density $\kappa$ is calculated 
by $\kappa^{-1} = d^2E/dn^2$, where n is the density.

First consider the contribution to $E_{ring}$ from the diffusive part.
For $\ell >> s_0$, where $s_0 \equiv (2 \pi e^2 \nu / \epsilon)^{-1}$
is the screening length in the Thomas-Fermi approximation,  
the leading contribution to $E_{ring} $ is 
$\sim D/ \ell^4$.  This yields a non-singular correction to the
inverse-compressibility $ \left( \frac{1}{\kappa}
\right) \sim (k_F \ell )^{-3}$, which decreases the compressibility
with increasing disorder. For $\ell << s_0$, the contribution to 
($E_{ring}$) is $\sim D / s_0^4$. This is proportional to the density;
so it produces no correction to the compressibility.

Consider next the contribution of the ballistic part.  
This is similar to the classic calculation of
Gell-Mann and Brueckner\cite{Gell-Mann,Rajagopal}
and others, but with the lower cut-off in
the $q$-integral given by $\ell^{-1}$.  On evaluation, 
the energy per particle, to the leading order in disorder,
may be written in units of a Rydberg as

\begin{eqnarray}
{\cal E}_{ring} = {\cal E}^0_{ring} 
-A \: {s_0 \over \ell}
\label{Ering2}
\end{eqnarray}
Here ${\cal E}^0_{ring}$ is the contribution 
for $\ell \rightarrow \infty$, and the constant
$A \approx 4/\pi$. The disorder correction comes mostly
from a correction to the zero point energy of the plasmons. 
The corresponding additive correction to the inverse compressibility is

\begin{eqnarray}
\left( \frac{\kappa_o}{\kappa} \right)_{ring} \approx 
0.16 \: r^2_s ~\frac{s_0}{\ell}
\label{kappa.correction}
\end{eqnarray}
where $\kappa_o = \nu$ is the contribution to the compressibility
of the kinetic energy. 
Here we have taken $\tau$ to be independent of the density,
as is appropriate for the experimental systems in the 
immediate vicinity of the
metal-insulator transition\cite{Kravchenko3}.

The actual screening length $s$ is related to the compressibility through

\begin{eqnarray}
{s/s_0 = \kappa_0/\kappa}
\label{screening.length}
\end{eqnarray}
Thus the screening length increases as $\ell^{-1}$.
We look for the condition that $s>>\ell_0$, the value of $\ell$ 
at temperatures of the order of the Fermi-energy. 
Eq. (\ref{kappa.correction}) provides the dominant contribution
for $r_s>>1$ and gives the condition

\begin{eqnarray}
r_s \gtrsim 3 \: ( \omega_o \tau_0 )^{1/2}
\label{rs.critical}
\end{eqnarray}
where $\tau_0 $ is the scattering time and $\omega_o = \hbar / ma^2_o$. 

So far we have focused on the ring-diagram contribution to the ground state
energy. The ring diagrams take into account direct processes and are sufficient
for small momentum transfers even when $r_s$ is not small. For large
momentum transfers, processes beyond ring diagrams, representing 
exchange corrections at short distances, become important. We have calculated
the contributions from these additional processes following the
Hubbard interpolation scheme\cite{Hubbard}, in which the 
bare Coulomb interaction in the susceptibility is multiplied
by a factor $(1- F(q))$, where $F(q) = q^2 / 2(q^2 + k_F^2)$.
The leading order disorder correction to the ground state energy is 
essentially unchanged from that given in Eq. (\ref{Ering2}).

Before we proceed  further, a comment on the regime 
$\ell << s$ is in order. We are looking for  
the transition to an insulator in which 
$\kappa \sim s^{-1} \rightarrow 0$ while 
$\ell^{-1} \rightarrow \infty$.  This regime cannot be reached from
the opposite limit $\ell >> s$, in which we, in
common with others, can find no  singular correction to the
compressibility, and Finkelstein scaling holds.

We next consider the problem in a finite box of size $L$ much larger 
than $\ell(L)$ (or equivalently a temperature $T = DL^{-2}$). 
For our considerations to be meaningful 
it is necessary that $\ell(L) << s(L)$ for $L$
of order a few times $\ell$, i.e condition (\ref{rs.critical}),
be satisfied and remain consistently so
as $L$ is increased. To test the latter, we must first 
calculate the correction to $\ell$ as a function of $L$ through
the calculation of the conductance $g(L)$. 
In calculating  the corrections to the conductance
we assume that the condition
$s(L) >> L >> \ell(L)$ is satisfied and check later for its consistency.

In this limit, the bare Coulomb interaction appears 
in the exchange correction to the conductivity.
Consequently the infra-red singularity in $d = 2$ (and 3) 
is stronger than in the opposite limit.  
For $s << \ell$, the perturbative correction is proportional to 
$ln \: ( L/ \ell )$, with a universal (and negative) 
coefficient\cite{Altshuler}. The same processes with unscreened
Coulomb interactions give 

\begin{eqnarray}
\frac{\delta g}{g} \simeq - 
\frac{2^{1/2}}{\pi^2} \: r_s L/ \ell , 
\:\: for \: s >> L >> \ell \:\: .
\label{conductivity.correction}
\end{eqnarray}
whose coefficient is $r_s-$dependent.
This singular contribution arises 
from the contribution of momenta less than $\ell^{-1}$ and is related 
to the diffusion poles.

Next consider the Hartree corrections to the conductivity, which
tend to enhance the conductivity in the limit
$s << \ell$.  The interactions appearing in
Hartree-corrections do not depend on the total momentum of the
particle-hole pair carrying the current.
They involve characteristic momenta of $O (k_F )$.
The only effect of the interactions at 
small momentum transfers is to produce a 
$ln(s)$ enhancement to the triplet amplitude.
The Hartree terms provide the same
logarithmic corrections for $s >> \ell$ as in
the opposite limit, and so may be neglected compared to the
contribution of Eq. (\ref{conductivity.correction}).

We can now check for the consistency of the assumption of unscreened
interactions over the length scale $L$. The compressibility given by 
Eq. (\ref{kappa.correction}) is $L$ dependent through $\ell(L)$.
First we note that Eq. (\ref{conductivity.correction}) leads to a
linear decrease of the mean free path as $L$ is increased.
It introduces a length scale $L^*$ at which the mean free path
decreases to the atomic scale:

\begin{eqnarray}
L^* \approx \ell_0 ( 1 + {{\pi^2} \over {2^{1/2}r_s}} )
\label{L*}
\end{eqnarray}
The initial increase of the screening length, as $L$ is increased,
is given by

\begin{eqnarray}
s(L)  ~\approx ~ s(\ell_0) + \frac{\sqrt{2} r_s}{\pi^2}
\: \frac{s(\ell_0)}{\ell_0} \: ( L - \ell_0 )
\label{sL}
\end{eqnarray}
It is then easily seen that,
$s(L) >> L >>\ell(L)$ provided 
$\frac{s(\ell_0)}{\ell_0} >> \max ( 1 , \frac{\pi^2}{\sqrt{2} r_s })$.

We have also calculated the correction to 
the single particle density of states
for the case that $s >>\ell$,

\begin{eqnarray}
\frac{\delta \nu}{\nu} \simeq - 
\frac{1}{\sqrt{2} \pi} \: r_s L/ \ell.
\label{dos.correction}
\end{eqnarray}
Eq. (\ref{dos.correction}) implies that the leading correction
to the single particle density of states at zero 
energy is proportional to $- T^{-1/2}$. Eq. (\ref{dos.correction})
also implies that the single-particle self-energy is
momentum-dependent. The single-particle scattering 
time is then singularly modified, in a form similar to that
of the transport lifetime.

It ought to be stressed that the results of this paper only give the
leading high temperature corrections to the quantities calculated. 
However Eq. (\ref{conductivity.correction}) implies that the scale
for the low temperature phenomena is of the order of the Fermi-energy.
The leading correction suggests that provided the condition 
(\ref{rs.critical}) is fulfilled  the screening length 
is consistently much longer than the mean free path and the relevant 
size of the system $L$ so that the Coulomb interaction is unscreened 
inside L. The conductivity in that case rapidly tends to zero. 
In the opposite regime $s(\ell_0)<< \ell_0$, the singularities
found here are absent. 
The problem then is dominated by the 
diffusion processes
and the Finkelstein scaling equations,
which scale towards the metallic state are valid. 
In the transition regime, $s(L) \approx \ell(L)$, processes considered 
in both theories must be included. We hope to pursue such an analysis.
But since the initial flow downwards of the conductance in Eq. (\ref{L*})
is much faster than the
behavior in Finkelstein's theory, $s(\ell_0) \approx \ell_0$ 
may be taken as a good approximate condition for the metal-insulator
transition.

The insulating state with disorder and Coulomb interactions 
is most likely a glass exhibiting the Efros-Shklovskii\cite{Efros}
phenomena. The precise behavior in the critical regime of the transition 
to such a glassy state is a difficult question
which needs further study.

The major new result here is the demonstration of a route to a 
metal-insulator transition in two dimensions as density is decreased,
as is found in 
experiments\cite{Kravchenko1,Kravchenko2,Popovic,Hanien,Simmons,Coleridge}.
The metal-insulator transition is evident in the theory at
temperatures of $O (E_F)$, also as in experiments. 
The theory preserves Finkelstein's prediction that 
a magnetic field coupling to spins turns the metallic state insulating,
since $\gamma_t$ becomes irrelevant.
But the approach to the insulating state is likely to be different.

The most important prediction of the theory is the vanishing of
the compressibility as the transition is approached from the
metallic side.  Some existing observations\cite{Einstein} in
n-GaAs are consistent with the compressibility 
approaching zero as density is
decreased towards $r_s \approx 6.8$.  But the metal-insulator
transition was not monitored in this experiment.  We urge
simultaneous compressibility and transport measurements to check
Eq. (\ref{rs.critical}).  Frequency dependent transport and
susceptibility experiments are also suggested in the critical regime
to test the idea that the transition is to a glassy state.

Another prediction of the theory is the condition (\ref{rs.critical})
for the metal-insulator transition. 
There is not enough data to test this condition systematically.
What there is, is consistent with it in the $\tau$-dependence
and approximately in magnitude of the critical $r_s$. In the reported
results\cite{Kravchenko1,Kravchenko2,Popovic,Hanien,Simmons,Coleridge},
the metal-insulator transition occurs
at $\omega_0\tau_0$ of about $100$ and $r_s$ of about $20$ with $r_s$ 
at the transition showing slight increases as sample quality is improved.
Eq. (\ref{rs.critical}) also implies that $k_F \ell$ at the transition
point is of order unity, as is seen in the 
experiments\cite{Kravchenko1,Kravchenko2,Popovic,Hanien,Simmons,Coleridge}.

While considerations of the variation of compressibility in a
problem with Coulomb interactions lead to a metal-insulator transition
in two-dimensions, such a transition is already present in
Finkelstein's theory in three-dimensions.  However such
considerations change the nature of the transition.  We
urge a study of the variation in compressibility as well as a
study of frequency-dependent transport as well as magnetic 
susceptibilities near and across the metal-insulator
transition, in three dimensions as well.

The basic ideas of this paper are of interest to several other
electronic transitions, for example, the superconductor to
insulator transitions\cite{Goldman} and the quantum-Hall
transitions\cite{Wei}.

Q.Si would like to acknowledge the support of NSF Grant No. DMR-9712626,
a Robert A. Welch Foundation grant and an A. P. Sloan Fellowship. 
He would also like to thank Bell Labs and NHMFL/FSU for hospitality
during his visits and V. Dobrosavljevic for discussions.

\begin{figure}
\epsfxsize=5 in
\centerline{\epsffile{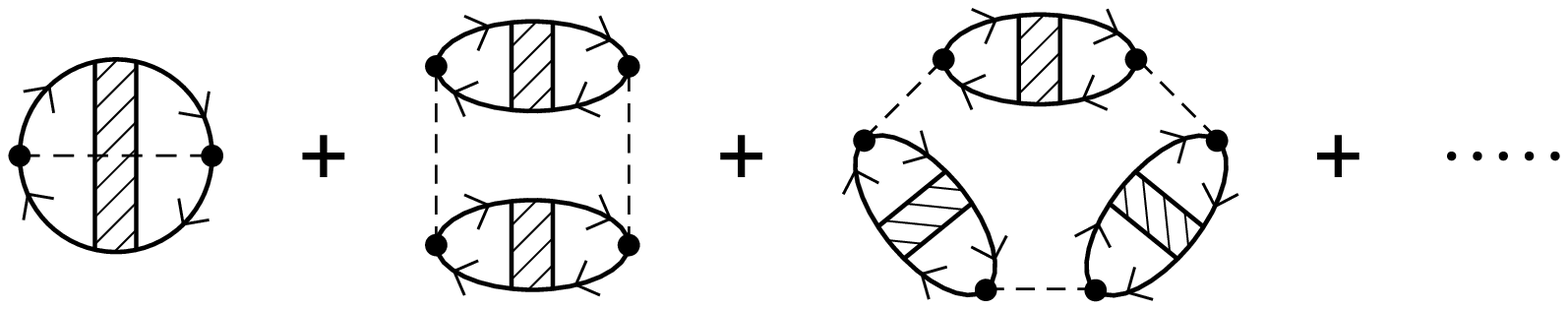}}
\caption{
Series for the ring contribution to the ground state energy with 
disorder. The hatched lines represent the t-matrix for impurity
scattering and the dashed lines represent the Coulomb interaction.}
\label{fig1}
\end{figure}


\begin{references}
\bibitem{impurity} The single-particle scattering rate is similar,
though not identical, to the transport scattering rate.
We have not made such a distinction in this paper.
\bibitem{Kravchenko3}
For carrier densities ($n$) close to the critical density
($n_c$), the high temperature conductivity 
$\sigma (n) \simeq \sigma (n_c) {\rm e}^{(n-n_c)/n_c} $
(S. V. Kravchenko, D. Simonian, and M. Sarachik, private
communication). This functional form implies that the mobility
is essentially $n-$independent near the metal-insulator
transition. Over a more extended density range the mobility
is a non-monotonic function of the carrier density \cite{Stern}.
\end{references}
\end{document}